\documentstyle[12pt,aaspp4]{article}
\begin{document}

\title{Local Approximations to the Gravitational Collapse of Cold Matter}

\author{Lam Hui and Edmund Bertschinger}
\affil{Department of Physics, Massachusetts Institute of Technology\\
Cambridge, MA 02139}

\begin{abstract}
We investigate three different local approximations for nonlinear
gravitational instability in the framework of cosmological Lagrangian
fluid dynamics of cold dust.  By local we mean that the evolution is described
by a set of ordinary differential equations in time for each mass element
with no coupling to other mass elements aside from those implied by the
initial conditions.  We first show that the Zel'dovich approximation
(ZA) can be cast in this form.  Next we consider extensions involving
the evolution of the Newtonian tidal tensor.  We show that two
approximations can be found that are exact for plane-parallel and
spherical perturbations.  The first one (``non-magnetic'' approximation,
or NMA) neglects the Newtonian counterpart of the magnetic part of
the Weyl tensor in the fluid frame and was investigated previously
by Bertschinger \& Jain (1994).  A new approximation (``local tidal'',
or LTA) involves neglecting still more terms in the tidal evolution
equation.  It is motivated by the analytic demonstration that it is
exact for any perturbations whose gravitational and velocity
equipotentials have the same constant shape with time.  Thus, the
LTA is exact for spherical, cylindrical, and plane-parallel
perturbations.  We tested all three local approximations in the case
of the collapse of a homogeneous triaxial ellipsoid, for which an exact
solution exists for an ellipsoid embedded in empty space and an
excellent approximation is known in the cosmological context.
We find that the LTA is significantly more accurate in general than
the ZA and the NMA.  Like the ZA, but unlike the NMA, the LTA
generically lead to pancake collapse.  For a randomly chosen mass
element in an Einstein-de Sitter universe, assuming a Gaussian random
field of initial density fluctuations, the LTA predicts that at least
78\% of initially underdense regions collapse owing to nonlinear
effects of shear and tides.
\end{abstract}


\section{Introduction}

The complexity of nonlinear gravitational instability challenges our
understanding of the universe.  Even though the law of gravity between
two bodies is very simple in the non-relativistic limit, the long-range
interactions among exceedingly many bodies leads to behavior that defies
simple analysis beyond the linear regime.  Computer simulation with
N-body methods provides a comprehensive approach to this problem, but
it suffers from finite dynamic range and computational expense.
Even more importantly, simulations do not increase our understanding
of dynamics without guidance from analytical approaches.

In this paper, we explore a class of what we call local approximations
for the nonlinear dynamics of self-gravitating cold matter.
By local we mean that each mass element behaves as if it evolves
independently of all the others once the initial conditions are specified.
This might sound quite implausible.  After all, mass elements do influence
each other through gravity. However, as we will demonstrate, the celebrated
Zel'dovich (1970) approximation (henceforth ZA) can be viewed as exactly
an approximation of this sort.

In the past several years, there have been various attempts to improve
upon the ZA, including the adhesion approximation (\cite{kps90}),
the frozen flow approximation (\cite{mlms92}), the frozen potential
approximation (\cite{bsv93}; \cite{bagpad94}), the truncated Zel'dovich
approximation (\cite{cms93}), and higher-order Lagrangian perturbation
theory (\cite{mbw95}) (note that the ZA can also be regarded as the
first-order solution in Lagrangian perturbation theory).  Most of them
are attempts to deal with the evolution of high density regions after
trajectories cross, when the ZA ceases to be adequate.   However, this
is a difficult problem.  Aside from the spherical model (\cite{peeb80})
and its cousins, there still exists little in the way of approximations
methods for post-collapse evolution.

In this paper, we will not try to tackle the problem of trajectory
crossing or the subsequent nonlinear evolution.  Instead we ask whether
one can improve upon the ZA even before orbits cross by seeking
generalizations of the ZA within the framework of local approximations.
In simple terms, a local approximation is one in which the evolution of
each mass element is described by a set of ordinary differential
equations in time in which there is no coupling to other mass elements,
aside from those implied by the initial conditions.  For instance, as
we will explain more fully later, the evolution of a given mass element
under the ZA is completely determined once the initial expansion,
vorticity, shear and density at this mass element are specified.
(The first three quantities correspond to the trace,
antisymmetric part and traceless symmetric part of the velocity
gradient tensor.)  The evolution of other mass elements have
no effect on the evolution of these quantities at this mass element.
In other words, under the ZA, all the information about other mass elements
is encoded in the initial conditions.  Once these are specified, each
mass element goes for its own ``free ride''!

We shall seek generalizations of the ZA by first systematically writing
down a set of Lagrangian evolution equations for the velocity and
gravity gradient for a given fluid element.  We discuss two local
approximations based on ignoring certain terms in the evolution equation
for the Newtonian tidal tensor.  One of them was introduced by Bertschinger
\& Jain (1994).  They used the fact that if a quantity known as the magnetic
part of the Weyl tensor vanishes in the Newtonian limit, the set of
exact Lagrangian fluid equations for cold dust becomes local.  (This
fact was proven in general relativity by \cite{mps93} following earlier
work of \cite{br89}; part of the motivation for such an assumption was the
statement of \cite{ellis71} that there is no counterpart to the magnetic
part of the Weyl tensor in Newtonian theory.)  Bertschinger \& Jain
then obtained the result that spindle (filamentary) collapse is favored
in general as opposed to pancake collapse. (Pancake collapse had been thought
--- correctly --- to be the generic outcome of gravitational collapse of
cold dust following the work of \cite{zel70}.)  Since then, it has been
shown that the magnetic part of the Weyl tensor does have a Newtonian
counterpart (\cite{bh94}; \cite{kp95} obtained equivalent results but
describe their conclusions slightly differently).  Until now, there has
been no quantitative determination of the magnetic Weyl term neglected
by Bertschinger \& Jain in the tidal evolution equation.  With reasonable
assumptions, this term may be negligible on super-horizon scales,
leading to ``silent universes'' (\cite{bmp95}).

Our second local approximation based on the tidal evolution equation is
entirely new.  It is based on dropping several more terms in addition
to the Weyl tensor term.  We will show why this is a better approximation
compared to the one proposed by Bertschinger \& Jain (1994).  In fact,
in tests this new approximation performs even better than the ZA,
both in cases where exact solutions are known and where numerical solutions
are calculated.  In this paper, we concentrate on a comparison of the
three local approximations for ellipsoids, with and without symmetries.

To understand the main ideas underlying these local approximation
methods, and how they differ from other approaches, it is useful
to draw an analogy with gravitational lensing.  Our use of Lagrangian
fluid equations is akin to solving the optical scalar equations
(\cite{sachs61}), whereby one follows the two-dimensional cross-section
of a congruence of light rays propagating through space.  Our approach
is similar, with light rays replaced by cold dust, and with the two-dimensional
cross-section replaced by the three-dimensional volume of a mass
element.  In fact, both approaches follow from the pioneering work
in general relativity by Ehlers (1961) and Kundt \& Tr\"umper (1961).
The first application of these methods to matter was by Hawking (1966),
who pioneered the covariant fluid approach to cosmological perturbation
theory.  The formalism was championed by Ellis (1971) and eventually
was applied to the formation of large scale structure (\cite{blh95} and
the references cited previously).  As in the case of gravitational
lensing, this approach can tell how a given (mass) element evolves
but does not give its trajectory.  The optical scalar equations do
not replace the gravitational lens equation, they supplement it.
Likewise, the local methods we advocate can supplement N-body simulations
or other approximations such as Lagrangian perturbation theory, by
providing accurate ways to follow the deformation of mass elements
as they evolve under gravity.

The organization of the paper is as follows. In \S \ref{za},
we show how the ZA is a local approximation. Section \ref{2loc}
presents two additional local approximations based on dropping
terms from the tidal evolution equation, and shows under what
circumstances these approximations are exact.  To compare the three
different local approximations for more general initial conditions,
in \S \ref{homell} we consider the motion of a homogeneous ellipsoid,
in both cosmological (Friedmann-Robertson-Walker background) and
noncosmological (vacuum) contexts.
The Weyl tensor and other relevant terms in the
tidal evolution equation are evaluated. In \S \ref{panspin} we discuss
how different nonlinear approximations predict pancake versus spindle
collapse from generic initial conditions, for which we also calculate
the collapse times.  Conclusions are presented in \S \ref{conclu}.
The Appendix presents some results of second-order perturbation theory.

\section{On the Zel'dovich Approximation}
\label{za}

In this section we review the Zel'dovich approximation starting from
the Eulerian fluid equations in comoving coordinates.  We then show
that it can be regarded as a local approximation.

The cosmological fluid equations for cold dust in a perturbed
Robertson-Walker universe with expansion scale factor $a(\tau)$ are
(\cite{blh95}):
\begin{equation}
  {\partial\delta\over\partial\tau}+\nabla_i\left[(1+\delta)v^i\,\right]=0\ ,
  \label{eulcont}
\end{equation}
\begin{equation}
  {\partial v^i\over\partial\tau}+v^j\nabla_jv^i=
    -{\dot a\over a}\,v^i-\nabla^i\phi\ ,
  \label{euler}
\end{equation}
\begin{equation}
  \nabla^2\phi=4\pi Ga^2\bar\rho\delta\ .
  \label{poisson}
\end{equation}
The mass density is $\rho = \bar\rho(\tau)(1+\delta)$ and $\vec v=
d{\vec x}/d\tau$ is the proper peculiar velocity where $\vec x$ is the
comoving spatial position and $\tau$ is the conformal time (hence,
$d\tau = dt/a$ where $t$ is the proper time).  We are neglecting
spatial curvature so that we can use Cartesian coordinates where
$\nabla^i=\nabla_i=\partial/\partial x^i$ for the $i$th spatial coordinate.

The trajectory of a fluid element is $x^i(\vec q,\tau)$ where $\vec q$
is a Lagrangian coordinate labeling the element, conventionally chosen
to be the initial position:
\begin{equation}
  x^i({\vec q},\tau) = q^i + \psi^i({\vec q},\tau)\ .
  \label{displacement}
\end{equation}
Now we introduce the Lagrangian time
derivative $d/d\tau \equiv\partial/\partial \tau + v^j\nabla_j$.
This time derivative commutes with $\partial /\partial q^i$.
Using $v^i=d\psi^i/d\tau$, we can rewrite equation (\ref{euler}) as
\begin{equation}
  {{d^2}\over d\tau^2}\psi^i+{\dot a\over a}{d\over d\tau}\psi^i
    -4\pi Ga^2{\bar\rho}\psi^i = - \nabla^i\phi - 4\pi Ga^2 {\bar\rho}
    \psi^i \ .
  \label{zeldovich}
\end{equation}
Each term on the left-hand side is first order in $\psi^i$.  The
right-hand side can be estimated from the Poisson equation (\ref{poisson}),
but first we need the mass density.  It follows in the Lagrangian
approach by noting that $\rho d^3x$ is conserved along a fluid streamline
provided $d^3x$ is computed from the mapping $\vec q\to\vec x$.
If there are no displacements, $\vec q=\vec x$ and $\rho=\bar\rho$.
The volume element follows from the Jacobian determinant, leading to
\begin{equation}
  \rho (\vec q,\tau) = {\bar \rho} \left\vert {\partial x^i\over
    \partial q^j}\right\vert^{-1}\ .
 \label{masscon}
\end{equation}
For small displacements the Jacobian may be expanded in a power series;
the first-order term gives
$\rho=\bar\rho(1-\partial\psi^i/\partial q^i)
+O(\psi^2$).  Now note that $\partial\psi^i/\partial x^i=(\partial\psi^i
/\partial q^j)(\partial q^j/\partial x^i)=\partial\psi^i/\partial q^i
+O(\psi^2)$.  Therefore, using equation (\ref{poisson}), we see that the
divergence of the right-hand side of equation (\ref{zeldovich}) vanishes
to first order in $\psi^i$.  If $\psi^i$ is longitudinal (i.e., has
vanishing curl), then the right-hand side itself vanishes to first order.
Displacements that grow by gravity are necessarily longitudinal in
linear theory.  The ZA consists of setting to zero the right-hand side
of equation (\ref{zeldovich}).  (It can be generalized to allow for a
transverse displacement; see \cite{buch93} and \cite{bs93}.)  Under the
ZA, the evolution of displacement thus obtained is used in equation
(\ref{masscon}) to get the density field.
The ZA is equivalent to first order Lagrangian perturbation theory
for the trajectories $\vec x\,(\vec q,\tau)$.

With vanishing right-hand side, equation (\ref{zeldovich}) is identical
to the linear perturbation evolution equation for $\delta$ (a fact that
becomes obvious when one notes $\delta=-\partial\psi^i/\partial q^i$ and
$d/d\tau=\partial/\partial\tau$ to first order in $\psi$).  This second-order
ordinary differential equation in time has two independent solutions that
we write $D_\pm(\tau)$ (\cite{peeb80}).  Taking the growing solution and
requiring $\psi^i$ to be longitudinal, we get the solution
\begin{equation}
  \psi^i({\vec q}, \tau) = D_+(\tau){\partial\varphi(\vec q\,)\over
    \partial q^i}
  \label{Zeld}
\end{equation}
where $\varphi(\vec q\,)$ is a displacement potential which is fixed
by initial conditions.

Next we will show that equations (\ref{displacement}) and (\ref{Zeld})
imply that the ZA displacement field is longitudinal in $x$-space
(the irrotational initial conditions already imply it is irrotational
in $q$-space), a first step needed before we show that the ZA is a local
approximation.  We have
\begin{eqnarray}
  \left(\vec\nabla\times\vec\psi\,\right)_i=
    D_+(\tau)\epsilon_{ijk}{\partial\over\partial x^j}
    \left({\partial\varphi\over\partial q^k}\right)=D_+
    \epsilon_{ijk}\left({\partial q^l\over\partial x^j}\right)
    \left({\partial^2\varphi\over\partial q^k\,\partial q^l}\right)\ ,
  \nonumber
\end{eqnarray}
where $\epsilon_{ijk}$ is the usual antisymmetric Levi-Civita symbol.
Now, note that the Jacobian matrix defined by the transformation of
equations (\ref{displacement}) and (\ref{Zeld}), $\partial x^j/\partial
q^l=\delta_{jl}+D_+\partial^2\varphi/\partial q^j\,\partial q^l$, is real
and symmetric.  By a theorem of linear algebra its inverse, $\partial
q^l/\partial x^j$, is also symmetric.  So is $\partial^2\varphi/\partial
q^k\,\partial q^l$ and, because they commute, so is their product.  Thus,
in the equation for $\vec\nabla\times\vec\psi$ above, $\epsilon_{ijk}$
is contracted with a matrix that is symmetric in $j$ and $k$, yielding
$\vec\nabla\times\vec\psi=0$ (\cite{zn83}).

The implication of this result is that $\vec\psi$ is longitudinal in
$\vec x$-space as well as in $\vec q$-space.  The same conclusions hold
for the velocity field $\vec v$, since it differs from $\vec\psi$ by
only a time-varying factor $\dot D_+/D_+$. As a result, under the
Zel'dovich approximation we can write
\begin{equation}
  \psi^i(\vec q(\vec x,\tau),\tau)=D_+(\tau){\partial\Phi(\vec x,\tau)
    \over\partial x^i}\quad\hbox{and}\quad v^i(\vec x,\tau)=\dot D_+(\tau)
    {\partial\Phi(\vec x,\tau)\over\partial x^i}\ ;\quad
    {d\Phi(\vec x,\tau)\over d\tau}=0\ .
  \label{KeyZeld}
\end{equation}
The last equation follows from the fact that $\partial\Phi/\partial x^i=
\partial\varphi/\partial q^i$ (cf. eq. \ref{Zeld}).  Recall that under the
ZA the right-hand side of equation (\ref{zeldovich}) vanishes.  Using
equation (\ref{KeyZeld}), we then get
\begin{equation}
  \vec v=-\dot D_+\left(4\pi Ga^2\bar\rho D_+\right)^{-1}\vec\nabla\phi=
    -{2\dot a \,f\over3\Omega_0 H_0^2}\vec\nabla\phi\ ,
  \label{velgrav}
\end{equation}
where $f\equiv d\ln D_+/d\ln a$.  Thus, in the Zel'dovich approximation,
the velocity field is always (not just to first order in $\vec\psi\,$)
proportional to the gravity field (\cite{kof91}).  It is clear geometrically
that this result must be correct for planar, cylindrical, or spherical flow
for growing mode initial conditions.  For plane-parallel flows, but not
otherwise, the coefficient of proportionality of the ZA is also correct,
so that the ZA is exact (until orbit-crossing) in one dimension.

We are now going to present the ZA from another point of view.  Similar
work has been done by Kofman \& Pogosyan (1995).  Our aim is to motivate
how one might improve the ZA by generalizing it to a broader class of
local approximations.  It will become clear shortly exactly what we mean
by local approximations.

Let us first give a brief summary of the Lagrangian fluid equations
(\cite{bj94}).   First of all, the gradient of the fluid velocity field
is decomposed into its trace, traceless symmetric and antisymmetric parts,
which are the expansion $\theta$, shear $\sigma_{ij}$ and vorticity
$\omega_{ij}$ respectively:
\begin{equation}
  \nabla_iv_j={1\over3}\,\theta\,\delta_{ij}+\sigma_{ij}+\omega_{ij}\,
  \quad \sigma_{ij}=\sigma_{ji}\ ,\quad
  \omega_{ij}=\epsilon_{ijk}\,\omega^k=-\omega_{ji}\ ,
  \label{gradv}
\end{equation}
where $2\vec\omega=\vec\nabla\times\vec v$.  Then, converting time
derivatives from Eulerian to Lagrangian, equation (\ref{eulcont}) becomes
\begin{equation}
  {d\delta\over d\tau}+(1+\delta)\,\theta=0
  \label{lagcont}
\end{equation}
Taking the trace of equation (\ref{euler}) and using equations
(\ref{poisson}) and (\ref{gradv}), one obtains the Raychaudhuri
equation:
\begin{equation}
  {d\theta\over d\tau}+{\dot a\over a}\,\theta+{1\over3}\,\theta^2+
    \sigma^{ij}\sigma_{ij}-2\omega^2=-4\pi Ga^2\bar\rho\delta\ ,
  \label{raych}
\end{equation}
where $\omega^2\equiv\omega^i\omega_i$. Similarly, taking the
antisymmetric and traceless symmetric parts of equation (\ref{euler})
gives respectively
\begin{equation}
  {d\omega^i\over d\tau}+{\dot a\over a}\,\omega^i+{2\over3}\,\theta\,
    \omega^i-\sigma^i_{\ j}\,\omega^j=0
  \label{vort}
\end{equation}
and
\begin{equation}
  {d\sigma_{ij}\over d\tau}+{\dot a\over a}\,\sigma_{ij}+{2\over3}\,
    \theta\,\sigma_{ij}+\sigma_{ik}\sigma^k_{\ j}+\omega_i\omega_j-
    {1\over3}\,\delta_{ij}\left(\sigma^{kl}\sigma_{kl}+\omega^2\right)=
    -E_{ij}\ ,
  \label{shear}
\end{equation}
where $E_{ij}\equiv\nabla_i\nabla_j\phi-(1/3)\,\delta_{ij}\,\nabla^2\phi$
is the gravitational tidal field.

In keeping with the spirit of Lagrangian fluid dynamics, we would like an
evolution equation for $E_{ij}$.  From equations (\ref{eulcont}) and
(\ref{poisson}), Bertschinger \& Hamilton (1994) derived
\begin{equation}
  {dE_{ij}\over d\tau}+{\dot a\over a}\,E_{ij}
    -\nabla_k\,\epsilon^{kl}_{\ \ \,(i}H_{j)l}+\theta E_{ij}
    +\delta_{ij}\,\sigma^{kl}E_{kl}-3\sigma^k_{\ \,(i}E_{j)k}
    -\omega^k_{\ \,(i}E_{j)k}
    =-4\pi Ga^2\rho\,\sigma_{ij}\ .
  \label{Eijdot1}
\end{equation}
Parentheses around a pair of subscripts indicates symmetrization,
e.g., $\sigma^k_{\ \,(i}E_{j)k}=(\sigma^k_{\ \,i}E_{jk} +
\sigma^k_{\ \,j}E_{ik})/2$.  The new quantity $H_{ij}$ is the Newtonian
limit of the magnetic part of the Weyl tensor in the fluid frame.
The definition and discussion of this term will be deferred until the
next section.

Equations (\ref{lagcont}) to (\ref{Eijdot1}) form a hierarchy of
Lagrangian fluid equations.  It is an incomplete set because we have
not stated the evolution equation for $H_{ij}$.  In order to arrive at
a local set, we must eliminate the gradient term in equation (\ref{Eijdot1}),
either by finding an approximation for $-\nabla_k\,\epsilon^{kl}_{\ \ \,
(i}H_{j)l}$ or by truncating the hierarchy in a way that eliminates our
need to determine it.

The ZA eliminates the need to calculate $H_{ij}$ by approximating the
evolution of the gravity field ---  equation (\ref{velgrav}) relates
$\vec\nabla\phi$ to $\vec v$.  As a result, the tidal tensor in the
ZA follows from the shear:
\begin{equation}
  E_{ij}=-{4\pi Ga\bar\rho\over Hf}\sigma_{ij}=
    -{3\Omega_0 H_0^2\over2\dot a\,f}\sigma_{ij}\ .
  \label{EequalSigma}
\end{equation}
Furthermore, the divergence of the gravity field is given in the ZA
by the velocity expansion scalar $\theta$ instead of the density
fluctuation.  Thus, the ZA is equivalent to solving the local
evolution equations
\begin{equation}
  {d\theta\over d\tau}+{\dot a\over a}\,\theta+{1\over3}\,\theta^2+
    \sigma^{ij}\sigma_{ij}-2\omega^2=
    {4\pi G a{\bar \rho}\over Hf}\theta\ ,
  \label{raych-zel}
\end{equation}
\begin{equation}
  {d\sigma_{ij}\over d\tau}+{\dot a\over a}\,\sigma_{ij}+{2\over3}\,
    \theta\,\sigma_{ij}+\sigma_{ik}\sigma^k_{\ j}+\omega_i\omega_j-
    {1\over3}\,\delta_{ij}\left(\sigma^{kl}\sigma_{kl}+\omega^2\right)=
    {4\pi G a{\bar \rho}\over Hf}\sigma_{ij}\ .
  \label{lagZeld}
\end{equation}
Together with equations (\ref{lagcont}) and (\ref{vort}), these give
a closed set of equations for the evolution of quantities for a single
mass element with no spatial gradients.  This is what we mean by locality.
Note that we have assumed the irrotational flow initial condition and so
$\omega_{ij}=0$ from equation (\ref{vort}) at all times before trajectories
intersect.  Equation (\ref{lagZeld}) can be written also as an evolution
equation for $E_{ij}$ by making use of equation (\ref{EequalSigma}) (\cite
{kp95}).  But it is clear that in terms of obtaining a closed set of local
equations, it is sufficient to stop at the level of the shear equation
(\ref{lagZeld}).

Hence, we have shown that the ZA is a local approximation based on
truncating the set of Lagrangian fluid equations at the shear
evolution equation by setting $E_{ij}$ proportional to $\sigma_{ij}$
and by approximating the gravitational source term in the Raychaudhuri
equation. It is then very natural to ask whether we can go further,
by using the exact Raychaudhuri equation and by truncating the system of
equations at the tidal evolution equation with a different approximation
from the ZA.

There is a simple argument for why we should expect to be able to
improve on the ZA.  It is well known that the ZA gives incorrect
results for spherical infall.  For spherical infall, the velocity
and gravity fields are isotropic around a point, so that $\sigma_{ij}=
E_{ij}=0$ at that point.  Yet, the ZA overestimates the collapse time
for a uniform spherical tophat.  The reason for this is that the ZA
does not obey the Poisson equation, so the right-hand side of equation
(\ref{raych-zel}) is not exact.  We can at least correct this term.
We have tested this approximation --- using equation (\ref{raych})
in place of equation (\ref{raych-zel}), and using equation ({\ref{lagZeld})
for the shear evolution --- and found that it works poorly aside from
spherical flow.  Thus, we seek improved approximations based on a more
accurate treatment of the tidal tensor.

\section{Two Local Approximations Based on the Tidal Evolution Equation}
\label{2loc}

As remarked in the last section, the hierarchy of Lagrangian fluid
equations can be truncated at the tidal evolution equation, provided
that we approximate, or eliminate, the $H_{ij}$ term (and possibly
other terms also).
If possible, we would like to find local approximations that retain the
successes of the Zel'dovich approximation.  These include giving the
correct results in linear perturbation theory and giving the exact
solution for plane-parallel flows.  Ideally, we would also like to
improve on the Zel'dovich approximation by giving exact results for
spherical and/or cylindrical flows.  We use these criteria in seeking
improved approximations.

Let's look at the magnetic part of the Weyl tensor more closely. The
definition is given in Bertschinger \& Hamilton (1994):
\begin{equation}
  H_{ij} \equiv -{1\over2}\,\nabla_{(i}H_{j)}-2\,v_k\,\epsilon^{kl}_
    {\ \ \,(i}E_{j)l}
  \label{defHij}
\end{equation}
where $H_i$ satisfies:
\begin{equation}
  \vec\nabla\times\vec H=-16\pi Ga^2\,\vec f_\perp\ ,\quad
  \vec\nabla\cdot\vec H=0\ .
  \label{Hfield}
\end{equation}
Here $\vec f_\perp$ is the transverse part of the mass current,
defined as follows:
\begin{equation}
  \vec f_\perp\equiv f-f_\parallel=\rho\vec v-f_\parallel\ ,\quad
  \vec f_\parallel = -{1\over 4\pi Ga^3}\vec
    \nabla\left(\partial\,a\phi\over\partial\tau\right) .
  \label{gcont}
\end{equation}
Using these definitions, we can rewrite equation (\ref{Eijdot1}) as
follows:
\begin{equation}
  {dE_{ij}\over d\tau} + {\dot a\over a}E_{ij} + M_{ij} =
     -4\pi Ga^2\rho\sigma_{ij}\ ,
  \label{Eijdot2}
\end{equation}
where
\begin{eqnarray}
  M_{ij} &\equiv&
    -\nabla_k\,\epsilon^{kl}_{\ \ \,(i}H_{j)l}+\theta E_{ij}
    +\delta_{ij}\,\sigma^{kl}E_{kl}-3\sigma^k_{\ \,(i}E_{j)k}
    -\omega^k_{\ \,(i}E_{j)k} \nonumber\\
    &=& -4\pi G a^2 \rho \nabla_{(i}v_{j)} - {1\over a}{d\over
      d\tau}(\nabla_i \nabla_j a \phi) \nonumber\\
    &=& -4\pi Ga^2\nabla_{(i}f_{\perp j)} -
      v_k\nabla^k\nabla_i\nabla_j\phi + v_{(i}\nabla_{j)}\nabla^2 \phi\ .
  \label{Kij}
\end{eqnarray}

Let us first consider plane-parallel flows, for which the ZA is exact.
The velocity and gravity gradient tensors may be written
\begin{equation}
  \nabla_i v_j=\theta\,\hbox{diag}(0,0,1)\ ,\quad
  \nabla_i\nabla_j\phi=\nabla^2\phi\,\hbox{diag}(0,0,1) \ ,
  \label{1-Dtensors}
\end{equation}
where $\hbox{diag}()$ denotes the elements of the diagonalized tensor.
Evaluating $M_{ij}$ using equations (\ref{Kij}), we find that the curl
$H_{jl}$ term as well as the sum of terms proportional to the tidal
tensor vanish identically.  The individual tidal terms do not vanish.
This result suggests two different closure schemes for the tidal evolution
equation (\ref{Eijdot2}).  The first one is to discard $\nabla_k\,
\epsilon^{kl}_{\ \ \,(i}H_{j)l}$.  The second is to discard the complete
tensor $M_{ij}$.  If some of the tidal terms of $M_{ij}$ were retained,
the resulting approximation would not be exact for one-dimensional
flows, hence would not improve on the Zel'dovich approximation.

The first choice, setting $H_{ij}=0$ in equation (\ref{Eijdot1}), was
proposed by Bertschinger \& Jain (1994):
\begin{equation}
  {dE_{ij}\over d\tau}+{\dot a\over a}\,E_{ij}
    +\theta E_{ij}
    +\delta_{ij}\,\sigma^{kl}E_{kl}-3\sigma^k_{\ \,(i}E_{j)k}
    -\omega^k_{\ \,(i}E_{j)k}
    =-4\pi Ga^2\rho\,\sigma_{ij}\ .
  \label{EijdotBJ}
\end{equation}
We shall call this the non-magnetic approximation (NMA).  Combined with
equations (\ref{lagcont})--(\ref{shear}), it provides a closed set of
local evolution equations.  The NMA was inspired, in part, by the remark
of Ellis (1971) that the magnetic part of the Weyl tensor has no Newtonian
counterpart.  However, it leads to unusual behavior, implying that cold dust
fluid elements generically collapse to spindles (\cite{bj94}).  Also,
Bertschinger \& Hamilton (1994) were able to derive equation (\ref{Eijdot1})
with $H_{ij}$ defined using equations (\ref{defHij})--(\ref{gcont}) from
Newton's laws in an expanding universe, as well as constraint and evolution
equations for $H_{ij}$ itself (the latter using post-Newtonian corrections),
from which we now know that $H_{ij}$ is not identically zero in the Newtonian
limit, aside from some special cases of high symmetry.

Thus, we are motivated to try the second approximation, setting
$M_{ij}=0$ in equation (\ref{Eijdot2}):
\begin{equation}
  {dE_{ij}\over d\tau} + {\dot a\over a}E_{ij} = -4\pi
    Ga^2\rho\sigma_{ij}\ .
  \label{new}
\end{equation}
Equation (\ref{new}) and equations (\ref{lagcont})--(\ref{shear}) form
our new set of closed local equations.  We shall call this the Local
Tidal Approximation (LTA) to distinguish it from equation (\ref{EijdotBJ}),
the non-magnetic approximation.

The LTA, like the ZA, is exact for plane-parallel flows prior to the
intersection of orbits.  What about spherically and cylindrically
symmetric flows, for which the ZA is not exact?  For the LTA,
we use equations (\ref{Kij}) to evaluate $M_{ij}$ for flows that
are spherically symmetrical around the fluid element under consideration.
As long as the gravity gradient is finite at the origin (a condition
that holds for any continuous finite-density mass distribution), this
restriction implies that all three eigenvalues of $\nabla_i\nabla_j
\phi$ are equal, so $E_{ij}=0$ identically (similarly $\sigma_{ij}=0$).
Equations (\ref{EijdotBJ}) and (\ref{new}) are satisfied trivially.
Thus, the LTA is exact for spherical mass elements.  So is the NMA.

Next we consider a non-singular fluid element on the symmetry axis of
a cylindrically symmetric flow.  By this we mean that two eigenvalues
of $\nabla_i\nabla_j \phi$ are equal and the third one vanishes and
similarly for $\nabla_i v_j$.  In this case we have
\begin{equation}
  \nabla_i v_j={1\over2}\,\theta\,\hbox{diag}(1,1,0)\ ,\quad
  \nabla_i\nabla_j\phi={1\over2}\,\nabla^2\phi\,\hbox{diag}(1,1,0) \ .
  \label{2-Dtensors}
\end{equation}
Using this, it is easy to show that the sum of tidal terms in the first
form of equation (\ref{Kij}) do not vanish, while, with equations
(\ref{poisson}) and (\ref{lagcont}),  the second form for $M_{ij}$
leads to $M_{ij}=0$.  Thus, the LTA is also exact for cylindrical
flows, while the NMA is not.  Bertschinger \& Jain (1994) erred in
saying that the NMA was exact for cylindrical flows.

One can generalize these results to show from the second form of
equation (\ref{Kij}) that $M_{ij}=0$ for any flow for which
$(\nabla^2\phi)^{-1}\nabla_i\nabla_j\phi$ equals $\theta^{-1}
\nabla_{(i}v_{j)}$ and is a constant tensor.  These conditions are
equivalent to saying that the orientation and axis ratios of the
gravitational and velocity equipotentials are constant for the mass
element under consideration.  Thus, the LTA is exact for flows with
equipotentials of constant shape.  Although this condition does not
always hold, it is valid for the growing mode in the linear regime
and it includes spherically and cylindrically symmetric flows as well
as plane-parallel flows.  Moreover, the gravitational potential
contours are more nearly spherical than the density contours around
a peak, so their shape would be expected to change relatively slowly
with time, suggesting that the LTA may be a good approximation in general.

In the linear regime, the LTA, NMA, and ZA all agree.  It is already
clear that they must differ in second-order perturbation theory; the Appendix
presents the calculation of $\nabla_k\,\epsilon^{kl}_{\ \ \,(i}H_{j)l}$
and $M_{ij}$.  However, it is more important to see how these various
approximations behave as collapse is approached.  We know already that
generic initial conditions lead to collapse along one dimension
(pancake) with the ZA (from \cite{bj94}) while the NMA leads to collapse
along two dimensions (spindle).  What about the LTA? How accurate is the
LTA for asymmetrical initial conditions?  Before answering these questions
we first examine the relative sizes of the terms in equation
(\ref{Eijdot2}) for an overdense homogeneous ellipsoid in an expanding
universe.

\section{Collapse of a Homogeneous Ellipsoid}
\label{homell}

We summarize here the equations of motion for an irrotational homogeneous
ellipsoid embedded in an expanding universe.
The various interesting quantities in the tidal evolution equation are then
calculated for the collapse of a particular ellipsoid.

We consider an irrotational homogeneous ellipsoid with proper axis lengths
$R_1$, $R_2$, and $R_3$ embedded in a Friedmann-Robertson-Walker background.
The equations of motion are (\cite{icke73}; \cite{ws79}):
\begin{equation}
  {d^2R_i\over dt^2}=-2\pi GR_i \left[{2\over3}\rho_b+\alpha_i
    (\rho_e-\rho_b)\right]\ ,
  \label{SW}
\end{equation}
where $t$ is the proper time ($dt=ad\tau$) and $\alpha_i$ is defined by
\begin{equation}
  \alpha_i = R_1 R_2 R_3 {\int^\infty_0} {ds\over ({R_i}^2 + s)
    {\sqrt{({R_1}^2+s)({R_2}^2+s)({R_3}^2+s)}}}\ .
  \label{Alpha}
\end{equation}
Here $\rho_e$ is the total density within the ellipsoid while $\rho_b$
is the density of the expanding universe surrounding the ellipsoid.
They are related to  the mean and perturbed densities used previously
by $\rho_b=\bar\rho$ and $\rho_e=\bar\rho(1+\delta)$.  We evaluate them
from the evolution of the axis lengths and the background expansion scale
factor:
\begin{equation}
  \rho_e R_1 R_2 R_3 = \rho_{eo}\ ,\quad
  \rho_b a^3 = \rho_{bo}\ ,
  \label{pseudomass}
\end{equation}
where $\rho_{eo}$ and $\rho_{bo}$ are constants.
Note that $\alpha_1+\alpha_2+\alpha_3=2$ and we assume $\Omega_0=1$.
Note also that since equation (\ref{SW}) is second order, there are in
general two independent modes.  We choose growing mode initial conditions.
For small $a$ the first order solution or, equivalently, the ZA result, is
\begin{equation}
  R_i(t)=a(t)X_i\left(1-\frac{1}{2}\alpha_{i0}\delta_0a\right)
  \label{axis2}
\end{equation}
where the $X_i$'s give the initial axis ratios, $\alpha_{i0}$ gives
the initial ellipsoid parameter, and $\delta_0$ gives the linear amplitude
of the density perturbation.  We set $\delta_0=1$ for overdense ellipsoids
without loss of generality.

Equation (\ref{pseudomass}) implies that the total mass,
including the mass inside the ellipsoid as well as outside, is
actually not conserved even though the mass inside the ellipsoid is.
Hence equation (\ref{SW}) can only be an approximation to the true
evolution of an initially homogeneous ellipsoid.  In general, one expects
that such an ellipsoid would cause the density of its immediate surroundings
to deviate from the cosmic mean.  Tidal fields from this perturbed external
material should then induce departure from homogeneity in the ellipsoid.
Based on results from an N-body simulation (S. D. M. White 1993, private
communication), we assume that it is a good approximation to ignore
departures from homogeneity inside and outside the ellipsoid when calculating
the evolution of the axis ratios.

It is noteworthy that equation (\ref{SW}) is exact if $\rho_b = 0$,
i.e., for a homogeneous ellipsoid in a vacuum.  Later in this section
we will test our approximations using the exact solution in this case.

The peculiar velocity field inside the homogeneous ellipsoid is described by:
\begin{equation}
  v_i = \left({\dot R_i\over R_i} - {\dot a\over a}\right)\,x_i
  \label{vellipse}
\end{equation}
and the gravitational potential within the ellipsoid is:
\begin{equation}
  \phi = \pi G a^2(\rho_e-\rho_b) \sum_i \alpha_i x_i^2\ .
  \label{phiellipse}
\end{equation}
(The $x_i$ are comoving coordinates and dots denote conformal time
derivatives.) Quantities like the expansion, shear, and tidal field
can be immediately read off from these expressions:
\begin{equation}
  \theta = \sum_i {\dot R_i \over R_i} - 3 {\dot a \over a}\ ,
  \label{thetaellipse}
\end{equation}
\begin{equation}
  \sigma_{ij} =\hbox{diag}\left({\dot R_i\over R_i} - {1\over 3}\sum_k {\dot
    R_k \over R_k} \right)\ ,
  \label{sigmaellipse}
\end{equation}
\begin{equation}
  E_{ij} = 2\pi Ga^2(\rho_e-\rho_b)\,\hbox{diag}\left(\alpha_i -
     {2\over3}\right)\ .
  \label{eijellipse}
\end{equation}
The tensor $M_{ij}$ defined in equation (\ref{Kij}) is given for the
homogeneous ellipsoid by
\begin{equation}
  M_{ij} = 2\pi G a^2 \rho_e\,\hbox{diag}\left[-2\sigma_{ij} +
    \left(\alpha_i - {2\over 3}\right)\theta - {\delta\over1+\delta}\,
    \dot\alpha_i \right]\ .
  \label{mijellipse}
\end{equation}

Using the time evolution of $R_i$ given by equation (\ref{SW}),
the evolution of the various quantities above can be calculated.
In particular, we are interested in the relative magnitude of various
terms in the tidal evolution equations (\ref{Eijdot1}) and
(\ref{Eijdot2}).  We integrated equations (\ref{SW}) and (\ref{Alpha})
numerically starting from equation (\ref{axis2}) at $a=10^{-8}$
with axis ratios $1 : 1.25 : 1.5$.  From the axis lengths $R_i$ and
their time derivatives, using equations
(\ref{thetaellipse})--(\ref{eijellipse})
we calculated the velocity and gravity gradient
terms inside the ellipsoid.  From these we then calculated the evolution
of $-\nabla_k\,\epsilon^{kl}_{\ \ \,(i}H_{j)l}$, $M_{ij}$, and
$4\pi Ga^2\rho\sigma_{ij}$ inside the ellipsoid.  Note that in this
test we do not integrate the tidal evolution equation itself; rather,
we evaluate the terms in it assuming that the system evolves according
to the homogeneous ellipsoid solution.  Although, as we noted above,
this solution is not exact, we are being self-consistent by evaluating
the various tensor quantities using equations
(\ref{thetaellipse})--(\ref{mijellipse}), which assume spatial homogeneity
inside the ellipsoid.

\begin{figure}[h]
  \figurenum{1}
  \begin{center}
  \plotone{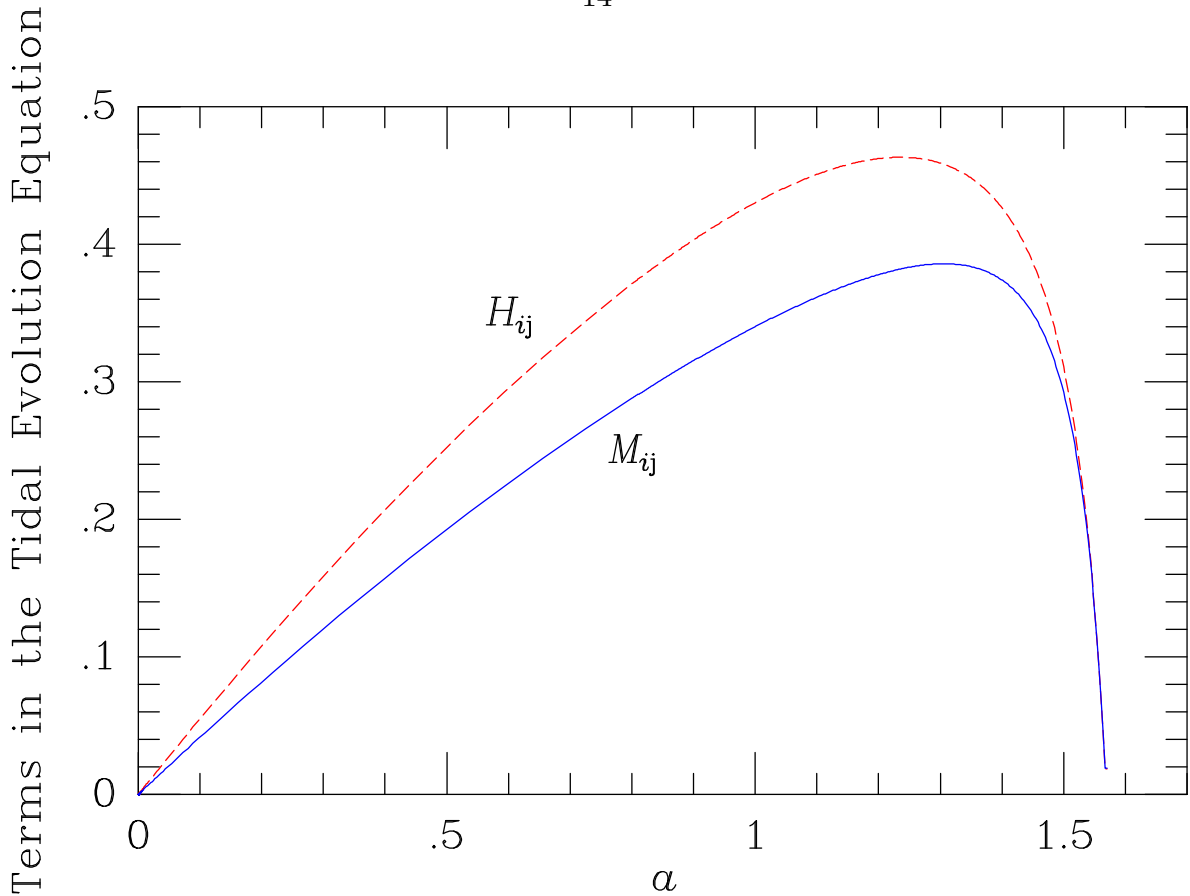}
  \caption{Evolution of the magnitudes of $-\nabla_k\,\epsilon^{kl}_{\ \
    \,(i}H_{j)l}$ (dashed line) and  $M_{ij}$ (solid line) divided by the
    magnitude of $4\pi Ga^2\rho\sigma_{ij}$, evaluated for a homogeneous
    ellipsoid with initial axis ratios $1:1.25:1.5$ embedded in an
    expanding universe.  The magnitude of a matrix is defined here as
    the square root of sum of squares of eigenvalues.}
  \end{center}
\end{figure}

Figure 1 shows the results of this calculation.  The magnitudes of
$-\nabla_k\,\epsilon^{kl}_{\ \ \,(i}H_{j)l}$ and $M_{ij}$ are divided
by the magnitude of $4\pi Ga^2\rho\sigma_{ij}$, where by the magnitude
of a matrix we mean the square root of the trace of its square.  We see
that $-\nabla_k\,\epsilon^{kl}_{\ \ \,(i}H_{j)l}$ and $M_{ij}$ are both
small compared to $4\pi Ga^2\rho\sigma_{ij}$ at both early and late times,
but not intermediate times (near maximum expansion).  Interestingly, the
magnetic term and $M_{ij}$ have similar magnitude throughout the collapse
process.

We can easily understand why $4\pi Ga^2\rho\sigma_{ij}$ is much bigger
than both $-\nabla_k\,\epsilon^{kl}_{\ \ \,(i}H_{j)l}$ and $M_{ij}$ at early
times using perturbation theory.  The shear is first order.  The last form
of equations (\ref{Kij}) is the best place to see that $M_{ij}$ is second
order: $f_{\perp i}$ is second order because, to first order, $f_i=\bar\rho
v_i$ is longitudinal (we assume irrotational initial conditions).  The other
contributions to $M_{ij}$ are obviously second order.  From the first form
of equation (\ref{Kij}), we conclude also that
$-\nabla_k\,\epsilon^{kl}_{\ \ \,(i}H_{j)l}$ is second order.
Expressions for these two tensors in second order perturbation theory
are given in the Appendix.  They are both nonzero in general.

The behavior of these quantities close to the moment of pancake
collapse can also be estimated analytically.  Suppose that the third
axis collapses while the other two axes still have finite lengths.
The term $4\pi Ga^2\rho\sigma_{ij}$ diverges at the moment of
pancake collapse because $\rho$ diverges and so does $\sigma_{ij}$,
owing to the $\dot R_3/R_3$ term in equation (\ref{sigmaellipse}).
For the behavior of $-\nabla_k\,\epsilon^{kl}_{\ \ \,(i}H_{j)l}$ and
$M_{ij}$ approaching collapse, we need to understand the behavior
of the $\alpha_i$'s.

It follows from equation (\ref{Alpha}) that $\alpha_1$ and $\alpha_2$
vanish in the limit of vanishing $R_3$ for finite $R_1$ and $R_2$,
because the integral is finite while the factor of $R_3$ in front vanishes.
Note also, by definition, the three $\alpha_i$'s always add up to $2$.
Hence $\alpha_3 = 2$ at collapse.  Moreover, it can be verified using
equation (\ref{Alpha}) that the $\dot\alpha_i$'s are finite at the moment
of collapse, assuming the $\dot R_i$'s are finite.  It can then be shown
using equations (\ref{thetaellipse})--(\ref{mijellipse}) that the particular
combination of $\alpha_i$'s conspires to render both $M_{ij}/4\pi
Ga^2\rho$ and $-\nabla_k\,\epsilon^{kl}_{\ \ \,(i}H_{j)l}/4\pi
Ga^2\rho$ finite.  Hence at the moment of pancake collapse,
$-\nabla_k\,\epsilon^{kl}_{\ \ \,(i}H_{j)l}$ and $M_{ij}$ are indeed
much smaller than $4\pi Ga^2\rho\sigma_{ij}$.

The fact that $-\nabla_k\,\epsilon^{kl}_{\ \ \,(i}H_{j)l}$ and
$M_{ij}$ are both small compared to $4\pi Ga^2\rho\sigma_{ij}$ at
early times and at the moment of pancake collapse suggests that the NMA
and LTA might both be good approximations.  However, Figure 1 shows
that these terms are not negligible throughout the collapse process.
Hence there is no guarantee that either approximation can reproduce
the correct  features of the collapse process.  In particular, we do
not know from these results whether the NMA or LTA would produce pancake
collapse given the initial conditions we have chosen.  We also do not
know which approximation will be more accurate for generic initial
conditions, although Figure 1 suggests that it may be better to
neglect $M_{ij}$ than $H_{ij}$.

\section{Pancakes Versus Spindles}
\label{panspin}

The oblate and prolate configurations are distinguished by the
signature of the eigenvalues of $E_{ij}$ and $\sigma_{ij}$.
For the collapsing oblate (pancake) configuration, the eigenvalues of
$E_{ij}$ have the signature $(-, -, +)$ and those of $\sigma_{ij}$ have
$(+, +, -)$.  For the collapsing prolate (spindle) configuration,
$E_{ij}$ has eigenvalues with signature $(-, +, +)$ and $\sigma_{ij}$
has $(+, -, -)$.  One way to see why this is true is by inspecting
equations (\ref{sigmaellipse}) and (\ref{eijellipse}).  For the
pancake configuration, one can use the fact that $\alpha_3$ is close
to $2$  (supposing collapse occurs in the third direction) while $\alpha_1$
and $\alpha_2$ almost vanish.  For the spindle configuration, suppose
that collapse occurs for the second and third direction and suppose
for simplicity that they collapse at the same rate.  Then from equation
(\ref{Alpha}), one can show that close to the spindle configuration,
$\alpha_3 \simeq \alpha_2 \simeq 1$ and $\alpha_1 \simeq 0$.  Using this
and equations (\ref{sigmaellipse}) and (\ref{eijellipse}), it is possible
to obtain the signature for the eigenvalues of $E_{ij}$ and $\sigma_{ij}$.
Note also that it is sufficient to consider only the divergent parts of
$E_{ij}$ and $\sigma_{ij}$ to get the right signatures.

Consider equation (\ref{EijdotBJ}). This is the tidal evolution
equation of the NMA, which ignores the magnetic part of the Weyl
tensor.  First of all, the term proportional to $\dot a/ a$ always
tends to decrease $E_{ij}$, encouraging spherical collapse.
But by the time the motion of the object under consideration
breaks away from the expansion of the universe, this term becomes
unimportant.  Suppose now that the object is close to the pancake
configuration with $E_{ij}$ having signature $(-, -, +)$ and
$\sigma_{ij}$ having $(+, +, -)$.  Then it can be seen that all
the terms favor pancake collapse (or favor neither pancakes nor
spindles) except the shear-tide coupling terms $\delta_{ij}\sigma^{kl}
E_{kl}-3\sigma^k_{\ \,(i}E_{j)k}$.  The net sign of these two terms
is such that the growth of $E_{ij}$ towards the pancake signature
is suppressed.  Suppose on the other hand that the object is close to the
spindle configuration with $E_{ij}$ and $\sigma_{ij}$ having signatures
$(-, +, +)$ and $(+, -, -)$ respectively.  Then all the terms, including
the shear-tide couplings, encourage the growth of tide towards the spindle
signature.  In other words, the NMA on the whole favors collapse toward
the prolate or spindle configuration.

Consider, on the other hand, the exact tidal evolution equation
(\ref{Eijdot2}).  For an object
with a very short third axis compared to the other two, we expect
$\alpha_3$ to be slightly less than but close to 2 and $\alpha_1$ and
$\alpha_2$ to be small and positive.  Substituting this into equation
(\ref{mijellipse}) and looking only at the most divergent terms, one can
verify that $M_{ij}$ has signature $(-, -, +)$ close to the pancake
configuration.  Using similar arguments, it can be deduced that $M_{ij}$
has signature $(-, +, +)$ close to the spindle configuration.  Hence,
$M_{ij}$ has the same signature as $E_{ij}$ close to collapse, whether
it be pancake or spindle; therefore it stabilizes collapse just like
the Hubble damping term proportional to $\dot a/a$.  Hence ignoring
$M_{ij}$, which is the LTA, does not favor spindles over pancakes.
This is very different from the NMA.

Equation (\ref{Kij}) tells us that $M_{ij}$ contains both
$-\nabla_k\,\epsilon^{kl}_{\ \ \,(i}H_{j)l}$ and the shear-tide
coupling terms.  We can now see what is wrong with the NMA --- for the
spindle configuration, $-\nabla_k\,\epsilon^{kl}_{\ \ \,(i}H_{j)l}$
has a signature that is opposite to the shear-tide coupling terms,
and it is large enough to reverse the spindle-enhancing effect of the
latter.  As a result, $M_{ij}$ as a whole, which includes the sum of
these terms, plays no favorites.

Numerical integration bears out this analysis.  We tested the LTA and
NMA by integrating the sets of local Lagrangian fluid equations which
are obtained by ignoring the relevant terms in the tidal evolution
equation: equations (\ref{lagcont}), (\ref{raych}), (\ref{shear}),
and either (\ref{EijdotBJ}) or (\ref{new}).  The tensor equations
were diagonalized along principal axes.  Initial conditions were
chosen using equations (\ref{axis2}) and
(\ref{thetaellipse})--(\ref{eijellipse}) so as to correspond to
the homogeneous ellipsoid model with initial axis ratios
$1 : 1.25 : 1.5$.  Given the numerical solution for $\theta(\tau)$
and $\sigma_{ij}(\tau)$, we then predicted the evolution of the
homogeneous ellipsoid axis lengths by solving equations
(\ref{thetaellipse})--(\ref{sigmaellipse}) for $\dot R_i/R_i$ and
numerically integrating it to get $R_i(\tau)$.  For comparison, we
also computed the prediction of the ZA for the axis evolution given
the same initial conditions.  We obtained the same results for the ZA
by integrating the local Lagrangian fluid equations (\ref{raych-zel})
and (\ref{lagZeld}) as we did from equation (\ref{axis2}).

\begin{figure}[h]
  \figurenum{2}
  \begin{center}
  \plotone{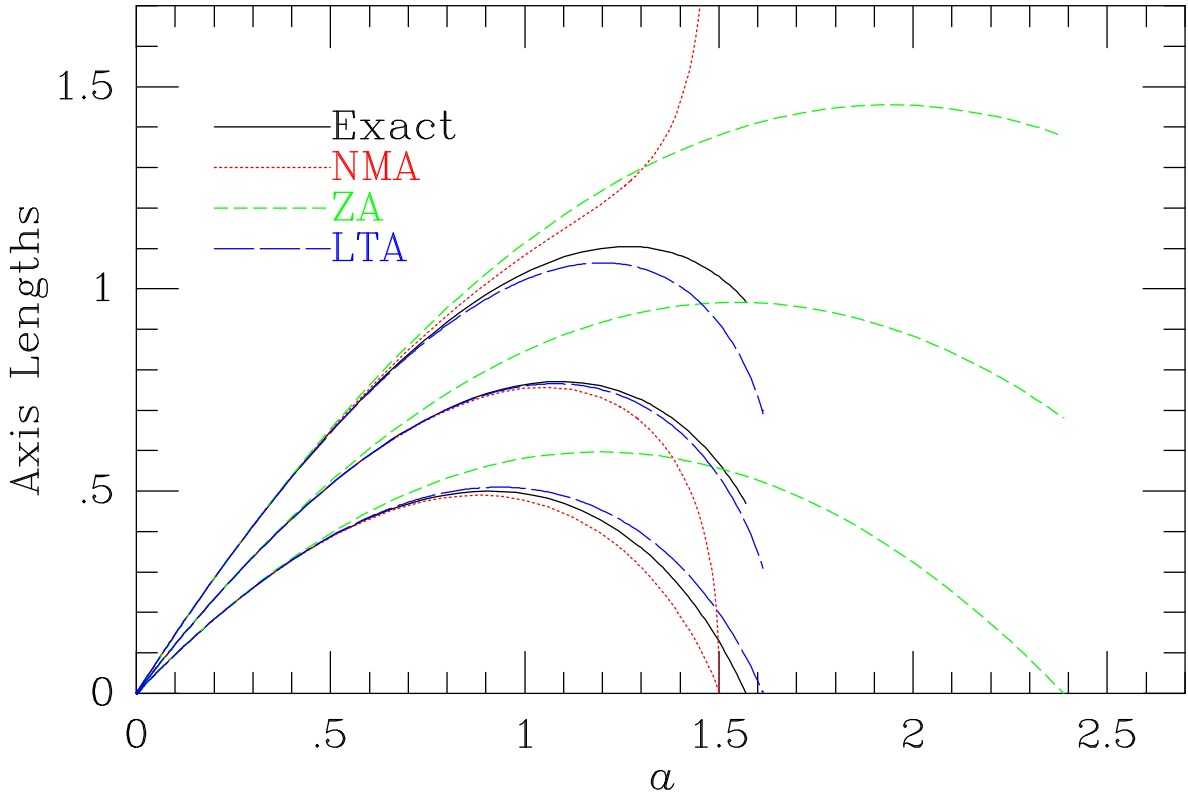}
  \caption{The evolution of axis lengths for a homogeneous ellipsoid
    embedded in an expanding universe.  The initial axis ratios are
    $1:1.25:1.5$.  The ``exact'' solution (ignoring development of
    inhomogeneity, solid curve) is compared with the ZA (short dashed
    curve) and two local approximations: the local tidal approximation
    (LTA, long dashed curve) and the non-magnetic approximation of
    Bertschinger \& Jain (NMA, dotted curve).}
  \end{center}
\end{figure}

Figure 2 compares the local approximations (ZA, LTA, NMA) for
the evolution of the axis lengths with each other and with the solution
given by integrating equations (\ref{SW}).  Both the ZA and LTA reproduce
the qualitative features of pancake collapse.  As we have already
noted, the NMA predicts collapse to a spindle instead of pancake.
For these initial conditions, at least, the LTA is even more accurate
than the ZA.  The LTA overestimates the expansion factor at collapse by
only 3\%, compared with 52\% for the ZA.  The LTA appears to rectify one
of the well-known problems with the ZA, namely the fact that it
underestimates the rapidity of collapse for non-planar perturbations.
This result is consistent with our observation in \S \ref{2loc} that
the LTA is exact for spherical and cylindrical symmetry.

\begin{figure}[h]
  \figurenum{3}
  \begin{center}
  \plotone{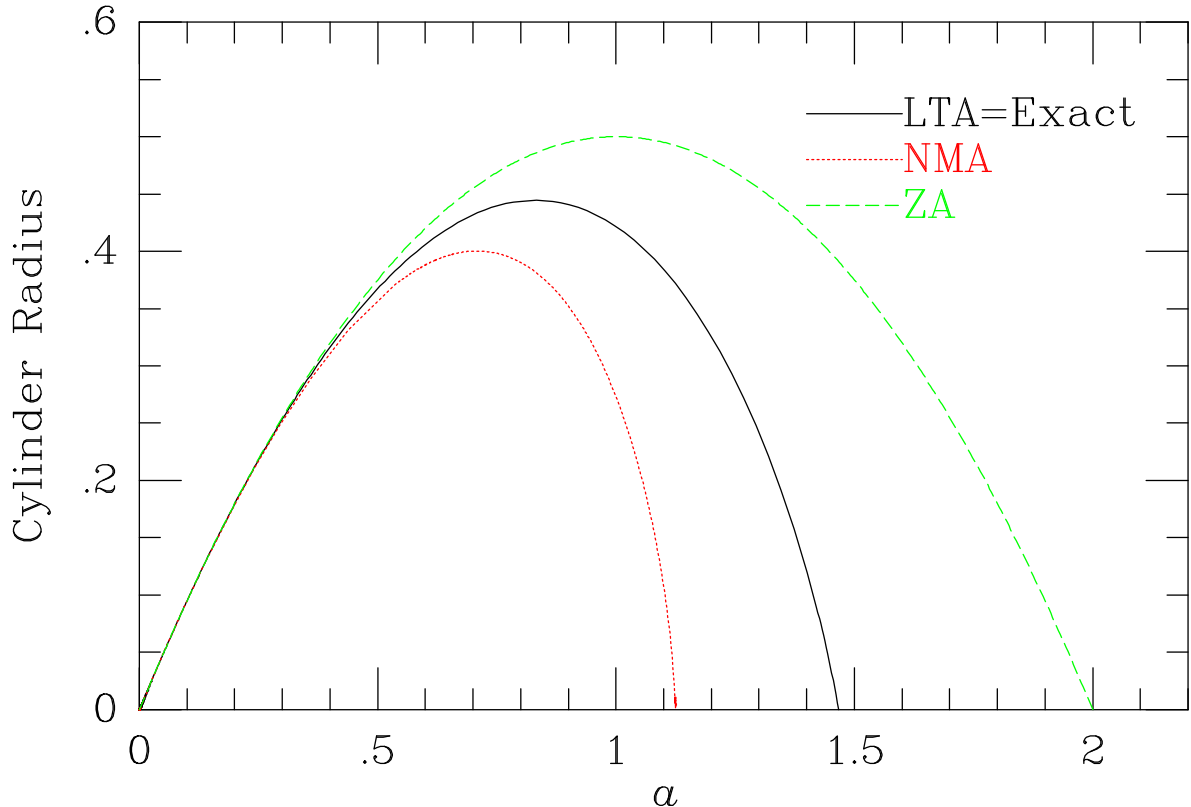}
  \caption{The evolution of the radius of a cylindrical perturbation
    in an expanding universe, corresponding to a homogeneous ellipsoid
    with axes $R:R:\infty$ (a cylinder).  The exact solution (solid curve)
    is compared with the ZA (short dashed curve) and the NMA (dotted
    curve).  The LTA is exact for this case.}
  \end{center}
\end{figure}

It is also useful to compare the local approximations with the exact
solution for cylindrically symmetric perturbations.  Consider a
homogeneous overdense cylinder in an Einstein-de Sitter universe,
with radius $R(t)$.  The equation of motion is given by Fillmore \&
Goldreich (1984).  It can be written in a form corresponding to equation
(\ref{SW}):
\begin{equation}
  {d^2R\over dt^2}=-2\pi GR \left[{2\over3}\rho_b+(\rho_c-\rho_b)
    \right]\ ,
  \label{cyl}
\end{equation}
where $\rho_c$ is the density inside the cylinder.  In fact, this
is identical to equation (\ref{SW}) for a homogeneous ellipsoid with
axis ratios $R : R : \infty$, for which $\alpha_1=\alpha_2=1$,
$\alpha_3=0$.  We repeated the comparison of local approximations
with the exact solution given by integrating equation (\ref{cyl}).
The results are shown in Figure 3.  The LTA is exact, while the NMA
underestimates the expansion factor at collapse (by 23\%) and the ZA
overestimates it (by 36\%).

\begin{figure}[h]
  \figurenum{4}
  \begin{center}
  \plotone{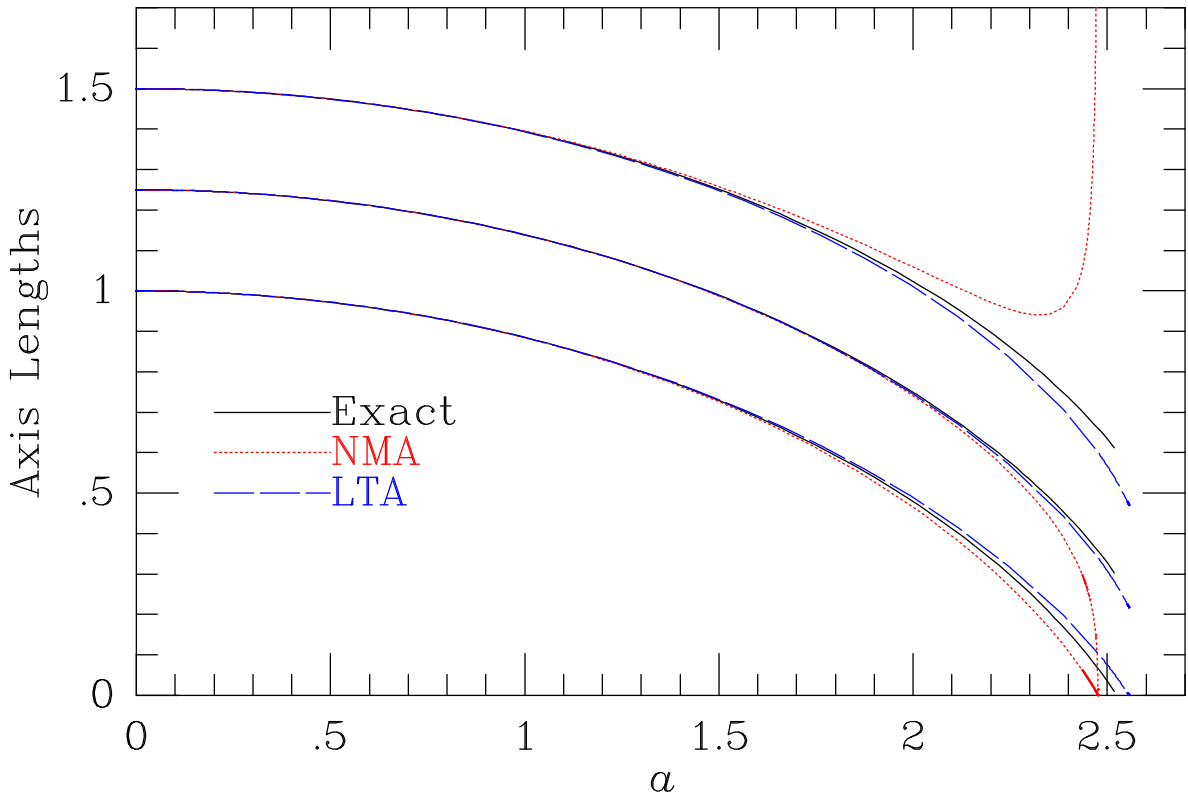}
  \caption{The evolution of axis lengths for a homogeneous ellipsoid
    embedded in empty space.  The initial axis ratios are $1:1.25:1.5$.
    The exact solution (solid curve) is compared with the predictions
    of LTA (long dashed curve) and NMA (dotted curve).}
  \end{center}
\end{figure}

An exact solution also exists for a homogeneous ellipsoid in a vacuum
(non-expanding) background (\cite{lin}).  It is easy to modify the
NMA and LTA equations for this case, by setting $a=1$ and $\rho_b=
\bar\rho=0$.  We did not integrate the non-cosmological analog of the
ZA.  As for Figure 2, we set the initial axis ratios to be $1:1.25:1.5$,
although in this case we set the initial velocity field to zero.
Figure 4 shows the results.  Once again we see that the LTA is rather
accurate for generic initial conditions (the collapse time here is 1.5\%
too large) and leads to pancake collapse, while the NMA incorrectly
predicts spindle collapse.

To compare the three local approximations (ZA, NMA, LTA) with more general
initial conditions, we follow the notations of Bertschinger \& Jain (1994)
and write traceless symmetric tensors in terms of a magnitude and an
angle:
\begin{equation}
  \sigma_{ij}={2\over3}\,\sigma\,Q_{ij}(\alpha)\ ,\quad
  E_{ij}={8\pi\over3}\,G\bar\rho a^2\,\epsilon\,(1+\delta)\,
  Q_{ij}(\beta)\ .
  \label{stq}
\end{equation}
We have introduced new scalars $\sigma\le0$, $\epsilon\ge0$,
$\alpha$ and $\beta$ ($0\le\alpha,\beta\le\pi$), and a one-parameter
traceless quadrupole matrix
\begin{equation}
  Q_{ij}(\alpha)\equiv\hbox{diag}\left[
  \cos\left(\alpha+2\pi\over3\right),\,\cos\left(\alpha-2\pi\over3\right),\,
  \cos\left(\alpha\over3\right)\,\right]\ .
  \label{quad1}
\end{equation}
With this parametrization, oblate configurations have $\cos\alpha>0$
while prolate configurations have $\cos\alpha<0$.  Of course, the
shape of a perturbation can change with time.  The equations of motion
for $\sigma$, $\epsilon$, $\alpha$, and $\beta$ for the NMA are given
by Bertschinger \& Jain.  For the LTA, their equations (13) and (14)
are changed to become
\begin{equation}
  {d\epsilon\over d\tau}-\theta\epsilon=-\sigma\cos\left(\alpha-\beta
    \over3\right)\ ,
  \label{epsdot}
\end{equation}
\begin{equation}
  {d\beta\over d\tau}=-{3\sigma\over\epsilon}\sin\left(\alpha-\beta
    \over3\right)\ .
  \label{betadot}
\end{equation}

One quantity of interest for general initial conditions is the
expansion factor at collapse, i.e., the linear overdensity when
a given mass element collapses.  Following Bertschinger \& Jain
(1994), we parametrize the initial conditions by $\epsilon_0$ and
$\alpha_0$, which are related to the values of $\epsilon$ and $\alpha$
in linear theory through $\epsilon=a\epsilon_0$ and $\alpha=\alpha_0$.
Because initially underdense perturbations can collapse if the shear is
sufficiently strong, we treat both initially overdense and underdense
perturbations by specifying $\delta_0=\pm1$, respectively ($\delta_0$
being related to $\delta$ in linear theory by $\delta=a\delta_0$).
The expansion factor at collapse, $a_c$, is determined by integrating
the local evolution equations for the LTA and NMA.  For the ZA, it is
simpler to use equations (\ref{displacement}) and (\ref{KeyZeld}), noting
that collapse occurs when the determinant of $\partial x^i/\partial q_j$
vanishes.  With our parametrization of the initial velocity and gravity
gradient tensors, it follows that
\begin{equation}
  a_c={3\over\delta_0+2\epsilon_0\cos(\alpha_0/3)}\quad\hbox{(ZA)}\ .
  \label{acza}
\end{equation}
The collapse expansion factor $a_c$ is defined to be the absolute
value of the linear overdensity when a given mass element collapses
to infinite density.  For example, an overdense spherical perturbation
collapses when $a_c=1.686$, while a cylindrical perturbation collapses
when $a_c=1.466$ and a plane-parallel perturbation collapses when $a_c=1$.
Although there exists no exact solution for arbitrary initial conditions,
it is informative to compare all three methods.  Based on our previous
results we expect the LTA to be accurate to a few percent.

\begin{figure}[h]
  \figurenum{5a}
  \begin{center}
  \plotone{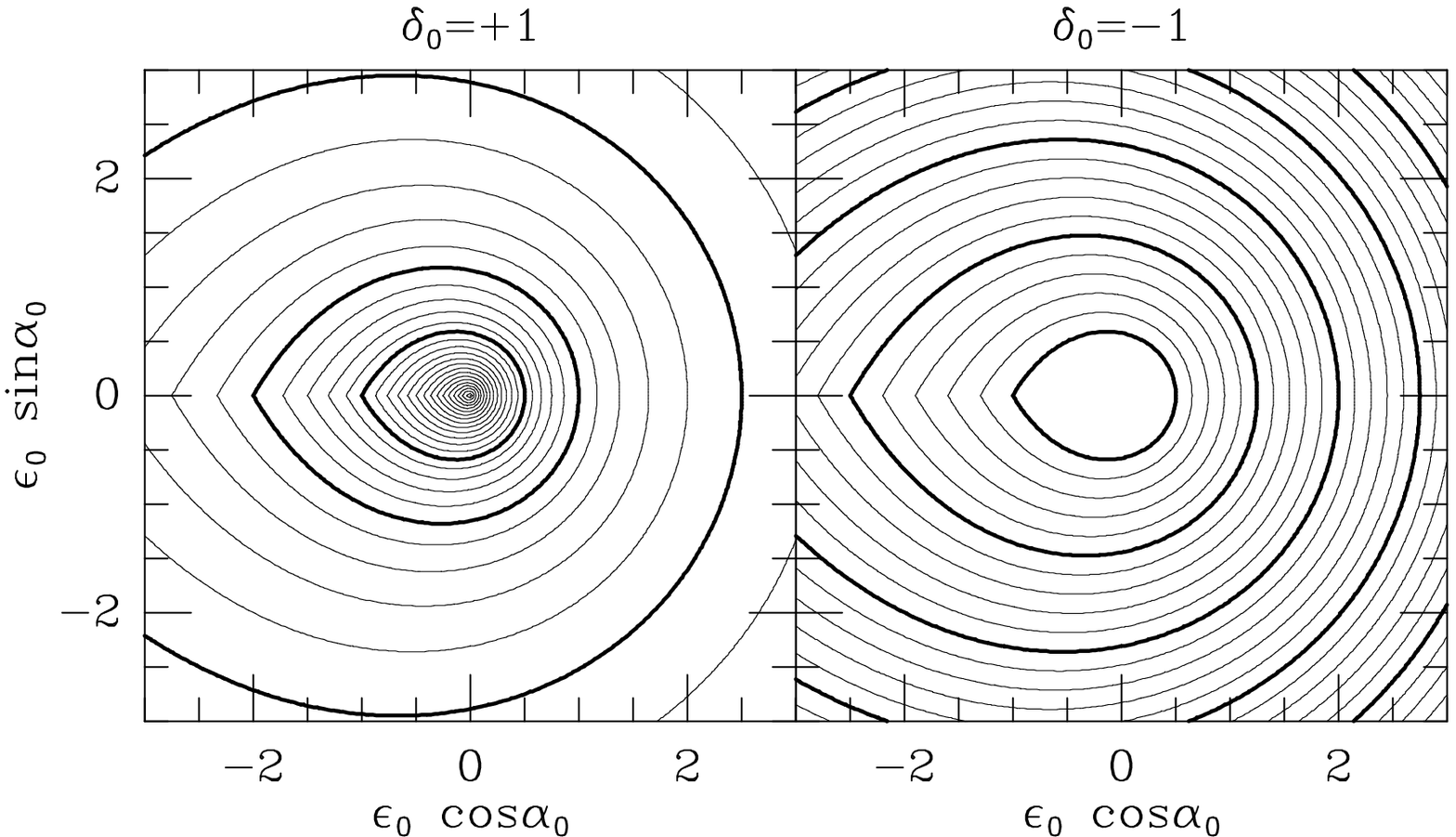}
  \caption{Contours of constant collapse time computed using the ZA,
    expressed by the cosmic expansion factor $a_c$ or its reciprocal,
    versus initial tidal field parameters.  Left panel: initial positive
    density perturbations.  The light (heavy) contours are spaced by 0.1
    (0.5) in $a_c$, with the outermost contour $a_c=0.4$ and the central
    value (corresponding to spherical collapse) $a_c=3.0$.  The ZA
    significantly overestimates the collapse time for low-shear perturbations.
    Right panel: initial negative density perturbations.  The light (heavy)
    contours are spaced by 0.1 (0.5) in $a_c^{-1}$, with the innermost
    contour $a_c^{-1}=0$ and the outermost one $a_c^{-1}=2.3$.  Initial
    perturbations in the central region do not collapse.  Perturbations
    are oblate (prolate) for $\epsilon_0\cos\alpha_0>0$
    ($\epsilon_0\cos\alpha_0<0$).}
  \end{center}
\end{figure}

\begin{figure}[h]
  \figurenum{5b}
  \begin{center}
  \plotone{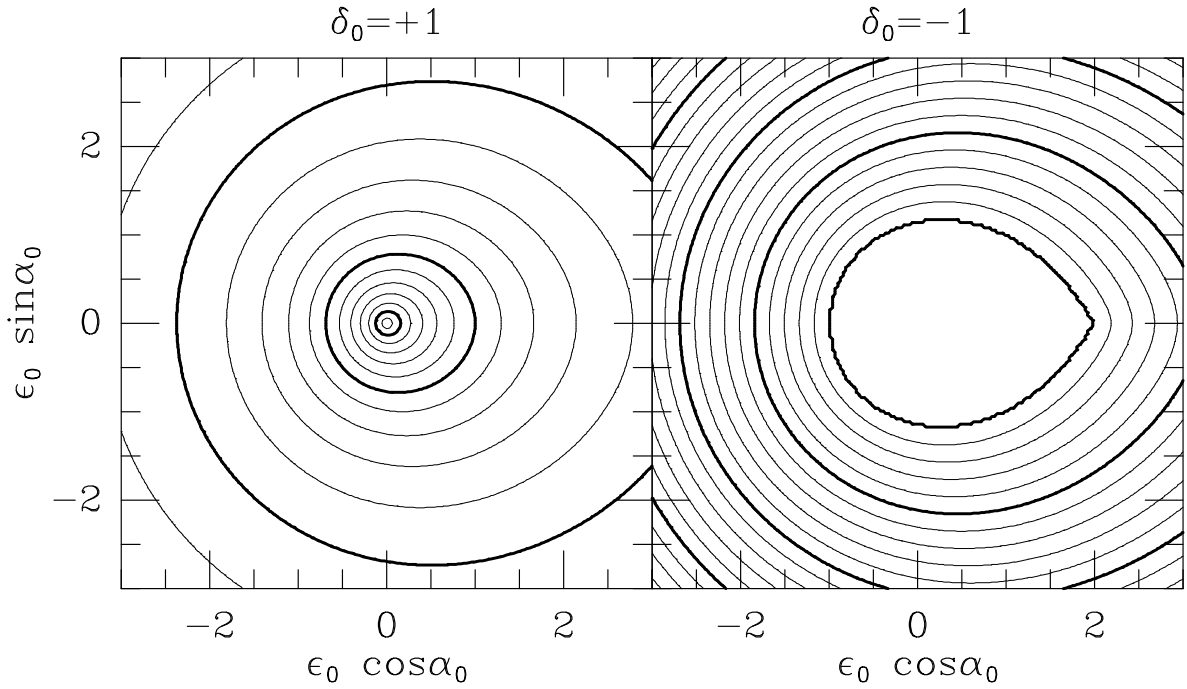}
  \caption{Same as Fig. 5a except that the NMA is used.  In the left
    panel the innermost contour is $a_c=1.6$.  In the right panel the
    outermost contour is $a_c^{-1}=1.8$.  The smaller extent of the
    contours for prolate configurations ($\epsilon_0\cos\alpha_0<0$)
    reflects the fact that the NMA favors prolate collapse.}
  \end{center}
\end{figure}

\begin{figure}[h]
  \figurenum{5c}
  \begin{center}
  \plotone{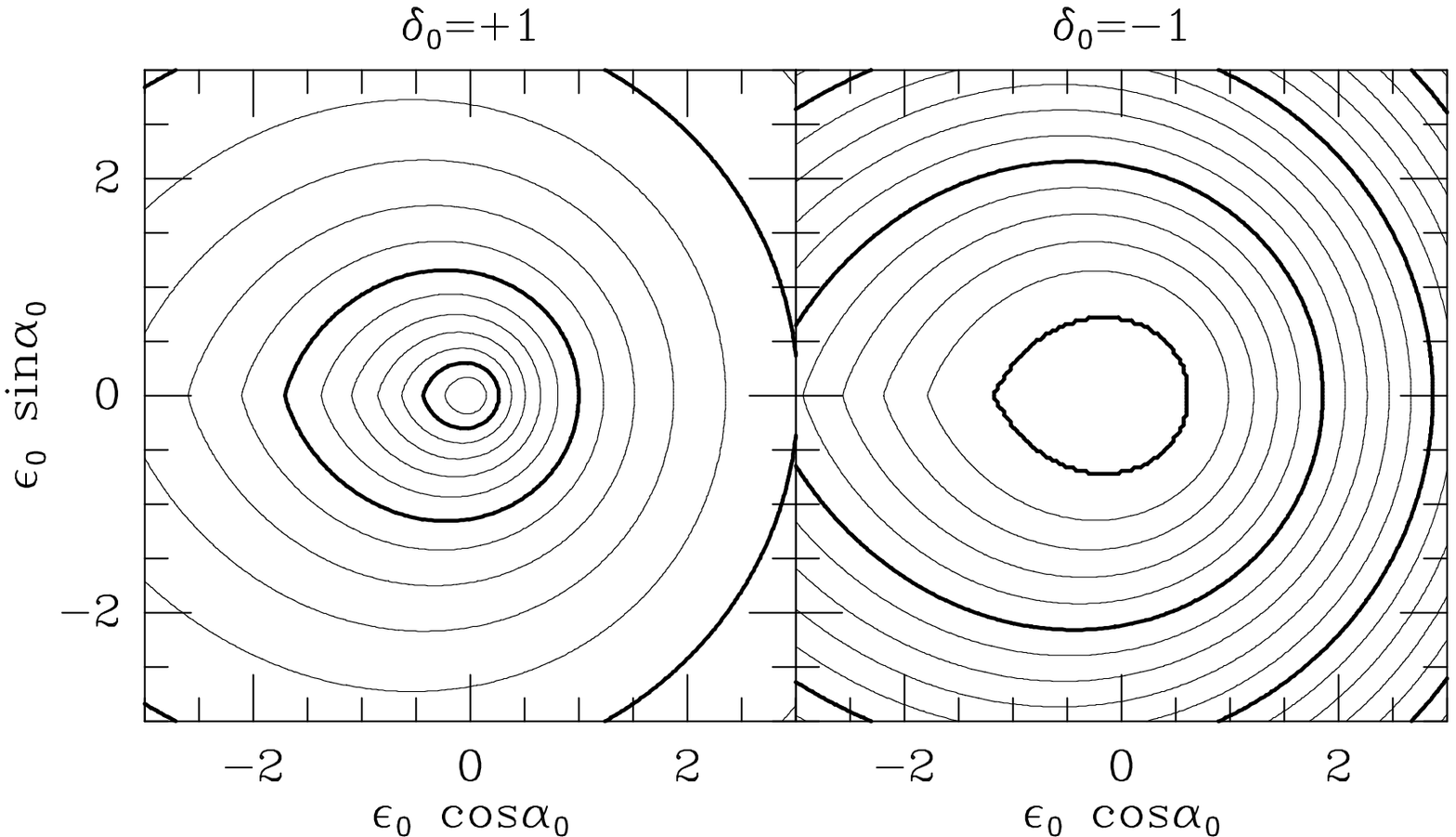}
  \caption{Same as Fig. 5a except that the LTA is used.  In the left
    panel the innermost contour is $a_c=1.6$.  In the right panel the
    outermost contour is $a_c^{-1}=1.6$.  The LTA, like the ZA, favors
    oblate (pancake) collapse over prolate (spindle) collapse.}
  \end{center}
\end{figure}

We plot contours of constant collapse time for different initial tidal
parameters $\epsilon_0$ and $\alpha_0$ for the three local
approximations in Figure 5.  In each part, the left panel gives results
for overdense perturbations while the right panel is for initially
underdense perturbations.  Figure 5b presents the same results as Figures
1 and 2 of Bertschinger \& Jain (1994).  We see that the LTA and ZA
are qualitatively similar, although the ZA overestimates the collapse
time for overdense configurations with small tide (near the center of
the figures).  According to the ZA, $a_c=3$ for spherical perturbations
while the exact value is $\frac{5}{3}(2/3\pi)^{2/3}=1.68647\ldots$.
Both the ZA and LTA indicate more rapid collapse for initially oblate
configurations.  As noted by Bertschinger \& Jain, initially prolate
configurations collapse faster in the NMA because according to its
incorrect dynamics initially oblate configurations must change shape
before collapsing to a spindle.

Bertschinger \& Jain (1994) also noted that shear can lead to collapse
of underdense perturbations.  From Figure 5, we see that that the size
of the non-collapsing region in parameter space (in the middle of the
right-hand panels) is largest for the NMA and smallest for the ZA,
indicating that the NMA underestimates the fraction of initial underdense
perturbations that can collapse, while the ZA overestimates it.  Using the
probability distribution of $\epsilon_0$ and $\alpha_0$ derived by
Bertschinger \& Jain for a Gaussian random field, we find that the
probability that a randomly chosen mass element will collapse is 0.780
for NMA, 0.888 for LTA, and 0.920 for ZA.  Thus, taking the LTA as the
most accurate, approximately 78\% ($=2\times0.888-1$) of the underdense
perturbations (and 100\% of the overdense ones) will collapse.  This
estimate neglects the crossing of mass elements, which increases the
likelihood of collapse by increasing the density. Indeed, we expect
every mass element collapses eventually in a perturbed self-gravitating
cold dust medium.

\section{Conclusion}
\label{conclu}

In this paper, we have discussed three different local
approximations for gravitational collapse of perturbations in an
expanding universe: the Zel'dovich approximation (ZA), the non-magnetic
approximation (NMA) of Bertschinger \& Jain (1994), and a new local
tidal approximation (LTA) introduced here.  Conventionally, the ZA
is presented as a mapping of Lagrangian to Eulerian positions.  However,
we showed that it can also be regarded as a certain truncation of
the set of Lagrangian fluid equations for the density, velocity gradient,
and tide following a fluid element of cold dust.  With the ZA, the
gravity gradient is explicitly proportional to the velocity gradient,
resulting in modifications to the Raychaudhuri and shear evolution
equations.  The tidal evolution equation need not be integrated in the
ZA because the gravity field acting on a mass element is given by a
simple extrapolation of initial conditions.  The other two
approximations we discuss extend the ZA by integrating the exact
Raychaudhuri and shear evolution equations, with approximations made
only to the tidal evolution equation.

All three local approximations are exact for plane-parallel
perturbations.  However, the behavior for other shapes of perturbations
shows significant differences in behavior.  The ZA is only approximate
for non-plane-parallel distributions.  The NMA is exact for spherical
perturbations but not cylindrical ones.  The LTA is exact for spherical
and cylindrical perturbations and, more generally, for any growing-mode
perturbations whose gravitational equipotential surfaces have constant
shape with time.

In order to test these approximations for non-symmetrical shapes, we
compared them in the case of the collapse of a homogeneous ellipsoid.
As expected from the results of Bertschinger \& Jain (1994), we find
that the NMA generically produces spindle-like singularities at collapse.
The LTA, on the other hand, generically produces pancakes, just like
the ZA.  For triaxial ellipsoids, we compared numerical integrations
of the local evolution equations with known solutions for a
homogeneous ellipsoid in both cosmological and vacuum backgrounds.
(An exact solution exists for the latter case while, in the former case,
the homogeneous ellipsoid solution is not really exact because tides will
cause the background, and then the ellipsoid itself, to become inhomogeneous.
However, these effects are expected to be small.)  We find that the LTA
is significantly more accurate than the ZA (see Fig. 2).

These results suggest we have found a promising new approximation for
nonlinear gravitational  instability.  However, we have only studied
the evolution of isolated irrotational perturbations.  Caution is needed
because we do not know how accurate the LTA is for more general initial
conditions, for example, those with vorticity.
Moreover, we do not know by how much the tide produced by other
mass elements degrades the accuracy.  External tides modulate the
equipotentials surrounding a mass element; qualitatively, we expect
little effect as long as the external tide evolves weakly or is small
compared with the trace part of the gravity gradient.  Quantitative
analysis is best done using N-body simulations, which we leave for later work.

The LTA has one significant limitation compared with the ZA.
It tells us only the internal state of a given mass element (density,
expansion rate, shear) and the tide on the element, but does not
give the position of the element.  However, for many purposes one cares
more about the internal evolution of a mass element than about its
position.  For example, simple models of galaxy formation are based
on spherical infall.  These can  be improved by inclusion of shear
and tides (\cite{bm93}; \cite{el95}).  Our approximations could lead to
even more accurate models of this sort.  Also, if one does need to know
the positions of mass elements, then one can always supplement the
Lagrangian fluid equations by the equation of motion for positions,
perhaps using the Zel'dovich approximation or higher-order Lagrangian
equations of motion.  In principle, by following the velocity gradient
for many mass elements, one can reconstruct the velocity field (up to an
irrelevant overall constant), and then integrate the positions with
$d\vec x/d\tau=\vec v$.  An equivalent procedure was suggested by
Matarrese et al. (1993).

Perhaps the most important reason for seeking new approximations like
the LTA is that we still lack a good understanding of the behavior
of collisionless systems under nonlinear gravitational instability.
Future work will tell whether local Lagrangian flow methods will
provide new insights.

\acknowledgements
We would like to thank Bhuvnesh Jain, Alan Guth, Rennan Bar-Kana, and
Jim Frederic for helpful discussions.  This work was supported by NASA
grant NAG5-2816.

\section*{Appendix: Second Order Calculation of $M_{ij}$}

We write the Eulerian density fluctuation field $\delta=\delta^{(1)}+
\delta^{(2)}+\ldots$ where $\delta^{(n)}$ is treated as being of
$n$th order in perturbation theory.  Similar expansions are used for
the velocity field and the scaled gravitational potential
\begin{equation}
  \hat\phi\equiv{\phi\over4\pi Ga^2\bar\rho}=-{1\over4\pi}\int d^3x'\,
    {\delta(\vec x^{\,\prime})\over\vert\vec x-\vec x^{\,\prime}\vert}\ .
  \label{sphi}
\end{equation}
For simplicity we shall assume an Einstein-de Sitter universe as in
the numerical examples presented in this paper.  In this case the
perturbation series is a series in $a(\tau)$.

Peebles (1980) presents the result for $\delta^{(2)}$, which we rewrite
using our variables as
\begin{equation}
  \delta^{(2)} = {5\over7} \left[\delta^{(1)}\right]^2 +
    \left[\vec\nabla\delta^{(1)}\right]\cdot\left[\vec\nabla\hat\phi^{(1)}
      \right]+{2\over7}F^2\ ,
  \label{delta2}
\end{equation}
where $F^2\equiv F^{ij}F_{ij}$ and
\begin{equation}
  F_{ij}\equiv\nabla_i\nabla_j\hat\phi^{(1)}\ ;
  \label{fij}
\end{equation}
note that $F^i_{\ \,i}=\delta^{(1)}$.  From the Euler equation (\ref{euler})
we get the first and second order terms of the peculiar velocity,
\begin{equation}
  \vec v^{\,(1)} = -{\dot a \over a} \vec\nabla\hat\phi^{(1)}\ ,\quad
  \vec v^{\,(2)} = -{2\over5}{\dot a\over a}\left[\vec\nabla\hat\phi^{(1)}
    \cdot\vec\nabla\,\right]\vec\nabla\hat\phi^{(1)}-{3\over5}{\dot a
    \over a}\vec\nabla\hat\phi^{(2)}\ ,
  \label{v12}
\end{equation}
where $\hat\phi^{(2)}$ is obtained using equation (\ref{sphi}) with
$\delta^{(2)}$.

We get $M_{ij}$ and $-\nabla_k\,\epsilon^{kl}_{\ \ \,(i}H_{j)l}$ from
equation (\ref{Kij}).  They vanish in first order; the second order results are
\begin{equation}
  M_{ij}=4\pi G\bar\rho a\dot a\left\{\delta^{(1)}F_{ij}-{7\over5}\nabla_i
    \nabla_j\hat\phi^{(2)}+{7\over5}\left[\vec\nabla\hat\phi^{(1)}\cdot
    \vec\nabla\,\right]F_{ij}+{2\over5}F^k_{\ \,(i}F_{j)k}\,\right\}\ ,
  \label{Mijsecond}
\end{equation}
\begin{equation}
  -\nabla_k\,\epsilon^{kl}_{\ \ \,(i}H_{j)l}=M_{ij}+4\pi G\bar\rho a\dot a
    \left\{3\delta^{(1)}F_{ij}-3F^k_{\ \,(i}F_{j)k}+\delta_{ij}\left(F^2-
    \left[\delta^{(1)}\right]^2\right)\right\}\ .
  \label{hijsecond}
\end{equation}
It can be verified that these quantities are traceless as expected using
equations (\ref{delta2}) and (\ref{fij}).  In general, neither vanishes
in second order perturbation theory.

\end{document}